\documentclass[conference]{IEEEtran}
\usepackage{cite}
\usepackage{amsmath,amssymb,amsfonts,amsthm}
\usepackage{graphicx}
\usepackage{textcomp}
\usepackage{xcolor}

\usepackage{algorithm,algorithmic}

\newtheorem{theorem}{Theorem}
\newtheorem{corollary}{Corollary}[theorem]

\newtheorem{example}{Example}

\begin{document}

\title{Backward Simulation for Sets of Trajectories}

\author{Yuxuan Xia$^{\circ}$, Lennart Svensson$^{\circ}$, {\'A}ngel F. Garc{\'\i}a-Fern{\'a}ndez$^{\star}$, Karl Granstr\"{o}m$^{\circ}$, 
Jason L. Williams$^{\dagger}$, \\
{\normalsize{}$^{\circ}$Dept. of Electrical Engineering, Chalmers
University of Technology, Sweden}\\
{\normalsize{}$^{\star}$Dept. of Electrical Engineering and Electronics,
University of Liverpool, United Kingdom}\\
$^{\dagger}${\normalsize{}Commonwealth Scientific and Industrial
Research Organisation, Australia}\\
{\normalsize{}Emails: firstname.lastname@chalmers.se, angel.garcia-fernandez@liverpool.ac.uk, 
jason.williams@data61.csiro.au}}

\maketitle

\begin{abstract}
This paper presents a solution for recovering full trajectory information, via the calculation of the posterior of the set of trajectories, from a sequence of multitarget (unlabelled) filtering densities and the multitarget dynamic model. Importantly, the proposed solution opens an avenue of trajectory estimation possibilities for multitarget filters that do not explicitly estimate trajectories. In this paper, we first derive a general multitrajectory forward-backward smoothing equation based on sets of trajectories and the random finite set framework. Then we show how to sample sets of trajectories using backward simulation when the multitarget filtering densities are multi-Bernoulli processes. The proposed approach is demonstrated in a simulation study.
\end{abstract}


\section{Introduction}

Multitarget tracking (MTT) refers to the problem of jointly estimating the number of targets and their trajectories from noisy sensor measurements \cite{mtt}. The major approaches to MTT include the joint probabilistic data association (JPDA) filter \cite{jpda}, the multiple hypothesis tracker (MHT) \cite{blackman2004multiple}, and random finite set (RFS) \cite{mahler2007statistical} based multitarget filters.

Vector-type MTT methods, e.g., the JPDA filter and the MHT, describe the multitarget states and measurements by random vectors. They explicitly estimate trajectories; i.e., they associate a state estimate with a previous state estimate or declare the appearance of a new target \cite{meyer2018message}. For multitarget filters based on set representation of the multitarget states, e.g., \cite{phd2,pmbmpoint}, and several of the particle filter based methods using the joint multitarget probability density (JMPD), e.g., \cite{kreucher2005multitarget,morelande2007bayesian}, time sequences of tracks cannot be constructed easily.

One approach to explicitly estimate trajectories is to add unique labels to the target states and estimate target states from the multitarget filtering density \cite{glmbconjugateprior,garcia2013two,aoki2016labeling}. This procedure can work well in some cases, but it may become problematic in challenging situations \cite{trackingbasedontrajectories,continuityPMBM}. A more advantageous approach to explicitly estimating trajectories for RFS-based multitarget filters is to generalize the concept of RFSs of targets to RFSs of trajectories \cite{trackingbasedontrajectories}. The set of trajectories posterior, which contains full information about the target trajectories, can be used to optimally estimate the set of trajectories \cite{trackingbasedontrajectories,continuityPMBM}. For multi-Bernoulli (MB) birth, this posterior may be labelled to consider sets of labelled trajectories, see \cite[Sec. IV.A]{trackingbasedontrajectories}, \cite{xia2019multi}, and also \cite{vo2019multi}.

Nevertheless, there are several MTT methods that can efficiently estimate the target states but that cannot easily produce trajectory estimation in a principled manner. For example: the set JPDA filter \cite{sjpda}, the variational MB filter \cite{williams2014efficient} and the JMPD based particle filter\footnote{Particle filter based methods usually suffer from history degeneracy.} \cite{kreucher2005multitarget}. Then an important research question arises: ``{\em can we leverage on filters that do not keep trajectory information to compute the posterior density of sets of trajectories?}''

In this paper, we show that this is true: the exact posterior of set of trajectories can be obtained from a sequence of multitarget filtering densities by using the multitarget dynamic model. Specifically, we derive a general multitrajectory forward-backward smoothing equation based on sets of trajectories. Contrary to existing literature on multitarget forward-backward smoothing \cite{vo2011closed}\cite[Chap. 14]{rfs}, the proposed backward smoothing recursion recovers the posterior over the set of trajectories, not simply the smoothed multi-target densities at each time step, which, even if labelled, may not be enough to provide trajectory information \cite[Ex. 2]{trackingbasedontrajectories}. Moreover, the proposed approach does not specify the form of the multitarget filtering density, thereby permitting the use of an arbitrary MTT method. This differentiates the proposed approach from multitarget forward-backward smoothers based on labelled RFSs \cite{beard2016generalised,liu2019computationally}, which cannot incorporate Poisson birth model in a theoretically sound manner and require that the multitarget filtering densities must be labelled.

As an application of the presented multitrajectory forward-backward smoothing equation, we show how sets of trajectories can be efficiently sampled from the smoothed multitrajectory density using backward simulation \cite{lindsten2013backward} when the multitarget filtering densities are MB processes \cite[p. 368]{mahler2007statistical}. The effectiveness of the proposed approach is validated in a simulation study.

The rest of the paper is organized as follows. The variable and density notations are introduced in Section II. In Section III, we present and derive the forward-backward smoothing equation for sets of trajectories. In Section IV, we present a tractable multitrajectory particle smoother using backward simulation and ranked assignments along with its linear Gaussian implementation. Simulation results are provided in Section V and conclusion is given in Section VI.

\section{Variables and Densities}

We briefly introduce the variables and densities used in this paper, see \cite{trackingbasedontrajectories} for more details. A trajectory is represented as $X=(t,x^{1:i})$ where $t$ is the initial time step of the trajectory, $i$ is its length and $x^{1:i} = (x^1,\dots,x^i)$ denotes a sequence of target states. Given a single target trajectory $X=(t,x^{1:i})$, the set of the target state at time $k$ is denoted $\tau_k(X)$. 

We are interested in the set of all trajectories that have passed through the surveillance area at some point in a given time interval. The set of trajectories limited in time interval $\alpha:\gamma$ is denoted ${\bf X}_{\alpha:\gamma}$. Given a set of trajectories, the set of target states at time $k$ is denoted ${\bf x}_k$ = $\tau_k({\bf X})$. A non-empty set ${\bf X}_{k:k}$ contains trajectories with initial time $k$ and length $1$, and therefore the set ${\bf x}_k$ of targets at time $k$ can be obtained as $\tau_k({\bf X}_{k:k})$. Also, given a set ${\bf x}_k$ of target states, we can construct its set of trajectories representation by changing the notation of target state from $x_k \in {\bf x}_k$ to $(k,x_k) \in {\bf X}_{k:k}$. Therefore, it holds that the multitarget density of ${\bf x}$ takes the same value as the multitrajectory density of ${\bf X}_{k:k}$, when evaluated for the corresponding set. Integrals for trajectories and sets of trajectories are defined in \cite[Eq. (3),(4)]{trackingbasedontrajectories}

We use $\delta_x(\cdot)$ and $\delta_x[\cdot]$ to represent the Dirac and Kronecker delta function centered at $x$, respectively. The multitarget Dirac delta function centered at ${\bf x}^\prime$ is denoted $\delta_{{\bf x}^\prime}({\bf x})$ \cite[Eq. (11.124)]{mahler2007statistical}, and is also valid for sets of trajectories. We use $p(\cdot)$ to denote the single target/trajectory density, $f(\cdot)$ to denote the multitarget filtering/prediction density, $g(\cdot|\cdot)$ to denote single target/trajectory transition density and $\pi(\cdot)$ to denote the multitrajectory density. Finally, we use ${\bf z}^k$ to denote the sequence of sets of measurements until time $k$.

\section{Forward-Backward Smoothing for Sets of Trajectories}

In this section, we present the forward-backward smoothing equations for sets of trajectories. Related proofs are given in the appendices. We consider the conventional assumptions for the dynamic model used in the RFS framework \cite[Sec. 13.2.4]{mahler2007statistical}. Given the current multitarget state ${\bf x}$, each target $x\in{\bf x}$ survives with probability $p_S(x)$ and moves to a new state with a transition probability $g(\cdot|x)$, or dies with probability $1-p_S(x)$.
The multitarget state at the next time step is the union of the surviving targets and new targets, which are born independently of the rest according to a Poisson point process with intensity $\lambda^b(\cdot)$.

We first present the multi-step prediction theorem for sets of trajectories and a resulting corollary that is important for the derivation of the forward-backward smoothing equations for sets of trajectories. Given $\pi({\bf X}_{\alpha:\eta}|{\bf z}^k)$, Theorem \ref{theorem:mt_predict} provides the $(\gamma-\eta)$-step predicted multitrajectory density $\pi({\bf X}_{\alpha:\gamma}|{\bf z}^{k})$\footnote{The measurements at each time step can also be represented using a vector.}. This is a generalization of the general prediction theorem for sets of trajectories \cite[Thm. 7]{trackingbasedontrajectories} to Poisson birth model and multi-step prediction. Note that it is also possible to derive the two-filter smoothing equation \cite{briers2010smoothing} for sets of trajectories using Theorem \ref{theorem:mt_predict}. 

\begin{theorem}\label{theorem:mt_predict}
    Given ${\bf X}_{\alpha:\gamma}$ with $\alpha\leq\eta<\gamma$ , $\eta\geq k$ and $\gamma\geq k+1$,  we define ${\bf W}^{\eta+1}$ as the set of trajectories that appeared after time $\eta$, ${\bf Y}^\eta$ as the set of trajectories present at time $\eta$ including the portions of trajectories before and after that time and ${\bf Z}^{\eta-1}$ as the set of trajectories present at a time before $\eta$ but not at $\eta$ such that ${\bf W}^{\eta+1}\uplus{\bf Y}^\eta\uplus{\bf Z}^{\eta-1}= {\bf X}_{\alpha:\gamma}$. We also consider ${\bf Y}_{\alpha:\eta}^\eta$ as the constrained ${\bf Y}^\eta$ in time interval $\alpha:\eta$, which implies that ${\bf Y}_{\alpha:\eta}^\eta\uplus{\bf Z}^{\eta-1}={\bf X}_{\alpha:\eta}$. Then the $(\gamma-\eta)$-step predicted multitrajectory density of $\pi( {\bf X}_{\alpha:\eta} |{\bf z}^{k})$ at time $\gamma$ is 
    \begin{multline}
        \label{eq:fb_sot_derivation_sub3}
        \pi({\bf X}_{\alpha:\gamma}|{\bf z}^{k}) = \prod_{\substack{(t,x^{1:i})\in{\bf Y}^\eta}}\Big( \left(1-p_S(x^i)+p_S(x^i)\delta_{\gamma-t+1}[i]\right) \\\times\prod_{j=\eta-t+1}^{i-1} g(x^{j+1} | x^j ) p_S(x^j) \Big)\pi( {\bf X}_{\alpha:\eta} |{\bf z}^{k})\pi_{\beta}({\bf W}^{\eta+1}),
    \end{multline}
    where $\pi_{\beta}(\mathbf{W}^{\eta+1})$ is the density of trajectories born at time $\eta+1$ and afterwards
    \begin{multline}\label{eq:multistep_birth}
        \pi_{\beta}(\mathbf{W}^{\eta+1}) = e^{-(\gamma-\eta)\int \lambda^b(x)dx}\prod_{(t,x^{1:i})\in{\bf W}^{\eta+1}}\lambda^b(x^1)\times\\\bigg( \left(1-p_S(x^i)+p_S(x^i)\delta_{\gamma-t+1}[i]\right) \prod_{j=1}^{i-1} g(x^{j+1} | x^j ) p_S(x^j) \bigg).
    \end{multline}
\end{theorem}
    
\begin{corollary}\label{corollary:mt_prediction}
    For $\gamma\geq k+1$, we have 
    \begin{equation}\label{eq:corollary1}
        \frac{\pi({\bf X}_{k:\gamma}|{\bf z}^{k})}{\pi({\bf X}_{k+1:\gamma}|{\bf z}^{k})} = \frac{\pi({\bf X}_{k:k+1}|{\bf z}^{k})}{f({\bf x}_{k+1}|{\bf z}^{k})}.
    \end{equation}
\end{corollary}

\begin{theorem}\label{theorem:FB_SoT}
    Given the multitarget densities $f({\bf x}_{k+1}|{\bf z}^{k})$, $f({\bf x}_{k}|{\bf z}^{k})$ and the multitrajectory density $\pi({\bf X}_{k+1:K} | {\bf z}^{K})$, the multitrajectory density in the time interval $k:K$ conditioned all the measurements until time $K$ is
    \begin{equation}\label{eq:fb_sot}
        \pi({\bf X}_{k:K} | {\bf z}^{K}) = \frac{ \pi({\bf X}_{k:k+1} | {\bf z}^{k}) \pi({\bf X}_{k+1:K} | {\bf z}^{K})  }{  f({\bf x}_{k+1} | {\bf z}^{k}) },
    \end{equation}
    where $\pi({\bf X}_{k:k+1} | {\bf z}^{k})$ is the predicted multitrajectory density obtained from $f({\bf x}_k|{\bf z}^{k})$.
\end{theorem}

Given the multitarget filtering densities computed in a forward recursion \cite{rfs}, Theorem 2 provides the backward recursion for sets of trajectories using the standard multitarget dynamic model with Poisson birth.\footnote{The presented backward recursions can also be adapted to MB (mixture) birth with minor modifications.} This equation is general and can be used to develop a range of different trajectory estimation algorithms. 

\begin{figure*}[!t]
    \footnotesize
    \hrulefill
    \begin{align}
        \pi\left(\left\{\left(t_1,x_1^{1:2}\right),\left(t_2,x_2^{1:2}\right)\right\}\right)&=\bigg({\cal N}\left([1,0]^{\mathrm{T}};[1,-1]^{\mathrm{T}},\mathrm{I}_2\right){\cal N}\left([2,0]^{\mathrm{T}};[2,1]^{\mathrm{T}},\mathrm{I}_2\right)\Big(\delta_{\left([2,-1]^{\mathrm{T}},[1,0]^{\mathrm{T}}\right)}\left(x_1^{1:2}\right)\delta_{\left([1,1]^{\mathrm{T}},[2,0]^{\mathrm{T}}\right)}\left(x_2^{1:2}\right)\notag\\&+\delta_{\left([2,-1]^{\mathrm{T}},[1,0]^{\mathrm{T}}\right)}\left(x_2^{1:2}\right)\delta_{\left([1,1]^{\mathrm{T}},[2,0]^{\mathrm{T}}\right)}\left(x_1^{1:2}\right)\Big)+{\cal N}\left([1,0]^{\mathrm{T}};[2,1]^{\mathrm{T}},\mathrm{I}_2\right){\cal N}\left([2,0]^{\mathrm{T}};[1,-1]^{\mathrm{T}},\mathrm{I}_2\right)\notag\\&\times\Big(\delta_{\left([1,1]^{\mathrm{T}},[1,0]^{\mathrm{T}}\right)}\left(x_1^{1:2}\right)\delta_{\left([2,-1]^{\mathrm{T}},[2,0]^{\mathrm{T}}\right)}\left(x_2^{1:2}\right)+\delta_{\left([1,1]^{\mathrm{T}},[1,0]^{\mathrm{T}}\right)}\left(x_2^{1:2}\right)\delta_{\left([2,-1]^{\mathrm{T}},[2,0]^{\mathrm{T}}\right)}\left(x_1^{1:2}\right)\Big)\bigg)\delta_1[t_1]\delta_1[t_2]\label{eq:example1}
    \end{align}
    \hrulefill
\end{figure*}

\begin{figure}[!t]
    \centerline{\includegraphics[width=\columnwidth]{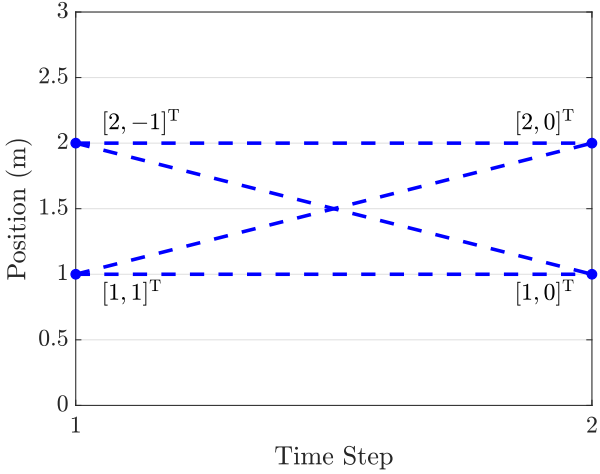}}
    \caption{One-dimensional scenario considered in Example \ref{example1}. At each time step, the target state is marked with $[\text{position},\text{velocity}]^{\mathrm{T}}$. Given the sets of target states at all times, there are 4 possible ways of constructing trajectories.}
    \label{example}
\end{figure}

\begin{example}\label{example1}
    Let us consider a two-dimensional two-target tracking scenario without target birth and death, illustrated in Fig. \ref{example}. We assume that the single target filtering density at each time step is a point mass represented by circles, and that targets move following a 1D constant velocity model with transition matrix $F=[1,1;0,1]$ and process noise covariance $Q=\mathrm{I}_2$, an identity matrix. The multitrajectory density $\pi({\bf X}_1:2|{\bf z}^2)$ can be recovered as \eqref{eq:example1} as there are four possible ways of linking the target states.
\end{example}

We proceed to give an explicit expression of $\pi({\bf X}_{k:k+1}|{\bf z}^{k})$ in \eqref{eq:fb_sot} using Theorem \ref{theorem:mt_predict}. Consider trajectories in time interval $k:k+1$, given a set ${\bf Y}$ of trajectories present at both time $k$ and $k+1$, a set ${\bf V}$ of trajectories present at time $k$ but not present at time $k+1$, and a set ${\bf B}$ of trajectories born at time $k+1$, we have that ${\bf X}_{k:k+1} = {\bf Y}\uplus{\bf V}\uplus{\bf B}$ and ${\bf x}_k = \tau_k({\bf Y}\uplus{\bf V})$. Given the multitarget filtering density $f({\bf x}_k|{\bf z}^{k})$, the predicted multitrajectory density is
\begin{multline}\label{eq:predicted_multitra}
    \pi({\bf X}_{k:k+1}|{\bf z}^{k}) = f({\bf x}_{k}|{\bf z}^{k})e^{-\int \lambda^b(x)dx}\prod_{(k+1,x^1)\in{\bf B}}\lambda^b(x^1)\\\times\prod_{(k,x^1)\in{\bf V}}
    \left( 1-p_S(x^1) \right) \prod_{(k,x^{1:2})\in{\bf Y}}\left( g(x^2|x^1)p_S(x^1) \right).
\end{multline}

\section{A Multitrajectory Particle Smoother}

In this section, we first present a multitrajectory particle smoother using backward simulation. Then we present a tractable implementation of the proposed method based on ranked assignments for MB filtering densities.

\subsection{Backward simulation for sets of trajectories}

A particle approximation of the multitrajectory density $\pi({\bf X})$ is 
\begin{equation}\label{eq:multiparticle_tra}
    \pi({\bf X})\approx \sum_{i=1}^\nu w^i \delta_{{\bf X}^i}({\bf X}),
\end{equation}
where $\nu$ is the number of particles and $w^i$ is the weight of the $i$th particle. We obtain $T$ particles $\{{\bf X}_{1:K}^i\}_{i=1}^T$ of the multitrajectory density $\pi({\bf X}_{1:K}|{\bf z}^K)$ with uniform weight $w^i=1/T$ by running backward simulation $T$ times for $k=K-1,\dots,1$. 

The basic idea of backward simulation is to make use of a particle filter to approximate the backward kernel that is used to generate samples from the joint smoothing density. In engineering literature, we often use the same notation for random variables and their realizations. Here we introduce a separate notation ${\bf X}^+_{k+1:K}$ for a realization of ${\bf X}^+_{k+1:K}$, whereas ${\bf X}_{k:K}$ denotes a realization of ${\bf X}_{k:K}$ and ${\bf X}_{k:k+1}$ is a part of ${\bf X}_{k:K}$. Given ${\bf X}^+_{k+1:K}$ and measurements ${\bf z}^{K}$, the backward kernel, in the context of sets of trajectories, is 
    \begin{equation}\label{eq:backward_kernel}
        \begin{split}
            \pi({\bf X}_{k:K}|{\bf X}^+_{k+1:K},{\bf z}^{K}) &= \frac{\pi({\bf X}_{k:K}|{\bf z}^{k})\pi({\bf X}^+_{k+1:K}|{\bf X}_{k:K})}{\pi({\bf X}^+_{k+1:K}|{\bf z}^{k})}\\
            &= \frac{\pi({\bf X}_{k:K}|{\bf z}^{k})\delta_{{\bf X}_{k+1:K}}({\bf X}^+_{k+1:K})}{\pi({\bf X}^+_{k+1:K}|{\bf z}^{k})}\\
            &= \frac{\pi({\bf X}_{k:k+1}|{\bf z}^{k})\delta_{{\bf X}_{k+1:K}}({\bf X}^+_{k+1:K})}{f({\bf x}^+_{k+1}|{\bf z}^{k})}\\  
            &\propto\pi({\bf X}_{k:k+1}|{\bf z}^{k})\delta_{{\bf X}^+_{k+1:K}}({\bf X}_{k+1:K}),
        \end{split}
    \end{equation}
where the first equality follows Bayes' rule and the conditional independence properties of state space models; in the second line we introduce the Dirac delta function; the third equality follows Corollary \ref{corollary:mt_prediction} and the the fact that the Dirac delta function is zero except when ${\bf X}_{k+1:K}={\bf X}^+_{k+1:K}$; and the last proportionality follows as $f({\bf x}^+_{k+1}|{\bf z}^{k})$ is a constant which does not depend on ${\bf X}_{k:K}$. It holds that $\tau_{k+1}({\bf X}^+_{k+1:K}) = \tau_{k+1}({\bf X}_{k:k+1})$ as they refer to the same set and conditioned on ${\bf X}^+_{k+1:K}$, and therefore $\tau_{k+1}({\bf X}_{k:k+1})$ is  deterministic. We elaborate on how to sample ${\bf X}_{k:K}$ from \eqref{eq:backward_kernel} in the following.

\begin{figure*}[!t]
    \footnotesize
    \hrulefill
    \begin{equation}\label{eq:tvb1}
        \pi({\bf Y} \uplus {\bf V} \uplus {\bf B}|{\bf z}^k) \propto \sum_{l_{1:n_k}}Q_{l_{1:n_k}}f_{l_{1:n_k}}\left(\left\{Y_1^|,\dots,Y_{n_y}^|,V_1^|,\dots,V_{n_v}^|\right\}\right)\delta_{n_k}[n_y+n_v]\prod_{i=1}^{n_b}\lambda^b\left(B_{i}^|\right)\prod_{i=1}^{n_v}\left( 1-p_S\left(V_i^|\right) \right)\prod_{i=1}^{n_y}\left( g\left(Y_i^{||}|Y_i^|\right)p_S\left(Y_i^|\right) \right)
    \end{equation}
    \hrulefill
    \begin{align}
        \begin{split}\label{eq:ud1}
            &\pi({\bf U}_{k:K}\uplus{\bf D}^+_{k+2:K}|{\bf S}_{k+1:K}\uplus{\bf D}_{k+2:K},{\bf z}^K) \propto \sum_{l_{1:n_k}}Q_{l_{1:n_k}} f_{l_{1:n_k}}\left(\left\{Y_1^{|},\dots,Y_{n_y}^|,V_1^|,\dots,V_{n_v}^|\right\}\right)\delta_{n_k}[n_y+n_v]\prod_{i=1}^{n_b}\lambda^b\left(B_{i}^|\right)\\&\quad\times\prod_{i=1}^{n_v}\left( 1-p_S\left(V_i^|\right) \right)\prod_{i=1}^{n_y}\left( g\left(Y_i^{||}|Y_i^|\right)p_S\left(Y_i^|\right) \right) \delta_{{\bf S}_{k+1:K}}({\bf U}_{k+1:K})\delta_{{\bf D}_{k+2:K}}({\bf D}^+_{k+2:K})
        \end{split}\\
        \begin{split}\label{eq:bwk}
            &\pi({\bf U}_{k:K}\uplus{\bf D}^+_{k+2:K}|{\bf S}_{k+1:K}\uplus{\bf D}_{k+2:K},{\bf z}^K) \propto \sum_{l_{1:n_k}}Q_{l_{1:n_k}} \sum_{\sigma_f\in\Sigma_{n_k}} \prod_{i=1}^{n_y}p_{l_{\sigma_f(i)}}(Y_i^|)\prod_{i=1}^{n_v}p_{l_{\sigma_f(i+n_y)}}(V_i^|) \delta_{n_k}[n_y+n_v]\prod_{i=1}^{n_b}\lambda^b\left(B_{i}^|\right)\\&\quad\times\prod_{i=1}^{n_v}\left( 1-p_S\left(V_i^|\right) \right)\prod_{i=1}^{n_y}\left( g\left(Y_i^{||}|Y_i^|\right)p_S\left(Y_i^|\right) \right) \sum_{\sigma_s\in\Sigma_{n_{k+1|K}}}\prod_{i=1}^{n_y}\delta_{X_{\sigma_s(i)}}(\bar{Y}_i)\prod_{i=1}^{n_b}\delta_{X_{\sigma_s(i+n_y)}}(\bar{B}_i)\delta_{n_{k+1|K}}[n_y+n_b] \delta_{{\bf D}_{k+2:K}}({\bf D}^+_{k+2:K})
        \end{split}\\
        \begin{split}\label{eq:bwk2}
            &\pi({\bf U}_{k:K}\uplus{\bf D}^+_{k+2:K}|{\bf S}_{k+1:K}\uplus{\bf D}_{k+2:K},{\bf z}^K) \propto \sum_{l_{1:n_k}}Q_{l_{1:n_k}} \sum_{\sigma_f\in\Sigma_{n_k}} \sum_{\sigma_s\in\Sigma_{n_{k+1|K}}}\prod_{i=1}^{n_b}\lambda^b\left(\bar{B}_{i}^|\right)\delta_{X_{\sigma_s(i+n_y)}}(\bar{B}_i) \\&\quad\times\prod_{i=1}^{n_y}p_{l_{\sigma_f(i)}}(Y_i^|)\left( g\left(\bar{Y}_i^{|}|Y_i^|\right)p_S\left(Y_i^|\right) \right)\delta_{X_{\sigma_s(i)}}(\bar{Y}_i)\prod_{i=1}^{n_v}p_{l_{\sigma_f(i+n_y)}}(V_i^|)\left( 1-p_S\left(V_i^|\right) \right)\delta_{n_k}[n_y+n_v]\delta_{n_{k+1|K}}[n_y+n_b] \delta_{{\bf D}_{k+2:K}}({\bf D}^+_{k+2:K})
        \end{split}\\
        \begin{split}\label{eq:bwk3}
            &\pi({\bf U}_{k:K}\uplus{\bf D}^+_{k+2:K}|{\bf S}_{k+1:K}\uplus{\bf D}_{k+2:K},{\bf z}^K) \propto \sum_{l_{1:n_k}}Q_{l_{1:n_k}} \sum_{{\cal A}^{ l_{1:n_k} }_{k}\in\mathbb{A}^{ l_{1:n_k} }_{k}} \prod_{(0,j)\in{\cal B}_k}\lambda^b\left(X_{(0,j)}^|\right)\delta_{X_{j}}(\bar{X}_{(0,j)}) \\&\quad\times\prod_{(\iota,j)\in{\cal Y}_k}p_{\iota}(X_{(\iota,j)}^|)\left( g\left(\bar{X}_{(\iota,j)}^{|}|X_{(\iota,j)}^|\right)p_S\left(X_{(\iota,j)}^|\right) \right)\delta_{X_{j}}(\bar{X}_{(\iota,j)})\prod_{(\iota,0)\in{\cal V}_k}p_{\iota}(X_{(\iota,0)}^|)\left( 1-p_S\left(X_{(\iota,0)}^|\right) \right)\delta_{{\bf D}_{k+2:K}}({\bf D}^+_{k+2:K})
        \end{split}
    \end{align}
    \hrulefill
        \begin{align}
            &\pi({\bf X}_{k:K}|{\bf X}^+_{k+1:K},{\bf z}^{K}) = \sum_{ l_{1:n_k} }Q_{l_{1:n_k}} \sum_{\mathcal{A}^{ l_{1:n_k} }_{k}\in\mathbb{A}^{ l_{1:n_k} }_{k}} \frac{\prod_{h\in{\cal A}^{ l_{1:n_k} }_{k}}w^h}{\sum_{ l_{1:n_k} }Q_{l_{1:n_k}} \sum_{\mathcal{A}^{ l_{1:n_k} }_{k}\in\mathbb{A}^{ l_{1:n_k} }_{k}}\prod_{h\in{\cal A}^{ l_{1:n_k} }_{k}}w^h}  p(X|h)\delta_{{\bf D}_{k+2:K}}({\bf D}^+_{k+2:K})\label{eq:mbm01_backward_kernel}\\
        \begin{split}
            &\pi({\bf X}_{k:K}|{\bf X}^+_{k+1:K},{\bf z}^{K}) \propto \sum_{ l_{1:n_k} }Q_{l_{1:n_k}}\sum_{\mathcal{A}^{ l_{1:n_k} }_{k}\in\mathbb{A}^{ l_{1:n_k} }_{k}} \prod_{h\in{\cal A}^{ l_{1:n_k} }_{k}}w^h  p(X|h)\delta_{{\bf X}^+_{k+1:K}}({{\bf X}}_{k+1:K})\\
            &\propto \sum_{ l_{1:n_k} }\prod_{i\in  l_{1:n_k} } r^{i} \prod_{i\in L_k\setminus  l_{1:n_k} }(1-r^{i})\sum_{\mathcal{A}^{ l_{1:n_k} }_{k}\in\mathbb{A}^{ l_{1:n_k} }_{k}} \prod_{h\in{\cal A}^{ l_{1:n_k} }_{k}}w^h  p(X|h)\delta_{{\bf X}^+_{k+1:K}}({{\bf X}}_{k+1:K})\\
            &= \sum_{ l_{1:n_k} }\prod_{i\in L_k\setminus  l_{1:n_k} }(1-r^{i})\sum_{{\cal A}^{ l_{1:n_k} }_{k}\in\mathbb{A}^{ l_{1:n_k} }_{k}}\prod_{(\iota,j)\in{\cal Y}_{k}} r^{\iota} w^{(\iota,j)}p(X|(\iota,j)) \prod_{(\iota,0)\in{\cal V}_{k}} r^{\iota}  w^{(\iota,0)}  p(X|(\iota,0))\prod_{(0,j)\in{\cal B}_{k}}w^{(0,j)}  p(X|(0,j))\delta_{{\bf X}^+_{k+1:K}}({{\bf X}}_{k+1:K})\\
            &\propto \sum_{ l_{1:n_k} }\sum_{{\cal A}^{ l_{1:n_k} }_{k}\in\mathbb{A}^{ l_{1:n_k} }_{k}}\prod_{i\in L_k\setminus  l_{1:n_k} }(1-r^{i})\prod_{(\iota,j)\in{\cal Y}_{k}} \frac{r^{\iota} w^{(\iota,j)}}{w^{(0,j)}}p(X|(\iota,j)) \prod_{(\iota,0)\in{\cal V}_{k}} r^{\iota}  w^{(\iota,0)}  p(X|(\iota,0))\prod_{(0,j)\in{\cal B}_{k}} p(X|(0,j))\delta_{{\bf X}^+_{k+1:K}}({{\bf X}}_{k+1:K})\label{eq:murty_weight}
        \end{split}
    \end{align}
    \hrulefill
\end{figure*}


\subsection{Backward simulation with multi-Bernoulli filtering}\label{section:multi-Bernoulli}

This section explains how to obtain samples of sets of trajectories using backward simulation when the filtering densities are MB processes \cite[p. 368]{mahler2007statistical}.
Suppose that the multitarget filtering density $f({\bf x}_k|{\bf z}^k)$ at time $k$ is an MB with $n_{k|k}$ Bernoulli components. Let $0 \leq r_1,\dots,r_{n_{k|k}} \leq 1$ be probabilities of existence and let $p_1(x),\dots,p_{n_{k|k}}(x)$ be existence-conditioned target state probability density functions. When ${\bf x}_k = \{x_1,\dots,x_{n_k}\}$ with $|{\bf x}_k|=n_k$, the MB process has a probability distribution of the ($\text{MBM}_{01}$) form \cite{pmbmpoint2}
\begin{equation}\label{eq:mbm01}
    f(\{x_1,\dots,x_{n_k}\}|{\bf z}^k) = \sum_{l_{1:n_k}}Q_{l_{1:n_k}}f_{l_{1:n_k}}(\{x_1,\dots,x_{n_k}\}),
\end{equation}
where
\begin{subequations}
   \begin{align}
    Q_{l_{1:n_k}} &\triangleq \prod_{i=1}^{n_{k|k}}\left( 1-r_i \right)\prod_{i=1}^{n_k}\frac{r_{l_i}}{1-r_{l_i}},\\
    f_{l_{1:n_k}}(\{x_1,\dots,x_{n_k}\}) &\triangleq \sum_{\sigma_f\in\Sigma_{n_k}}p_{l_{\sigma_f(1)}}(x_1)\dots p_{l_{\sigma_f(n_k)}}(x_{n_k}).\label{eq:mb01}
   \end{align} 
\end{subequations}
Here, $f_{l_{1:n_k}}(\{x_1,\dots,x_{n_k}\})$ denotes an $\text{MBM}_{01}$ specified by $l_{1:n_k}$ and $Q_{l_{1:n_k}}$ is the corresponding weight. Also, $l_{1:n_k}\triangleq(l_1,\dots,l_{n_k})$ and $\Sigma_{n_k}$ is the set that includes all the permutations of $(1,\dots,n_k)$. The summation is taken over all $l_1,\dots,l_{n_k}$ such that $1 \leq l_1<\dots<l_{n_k} \leq n_{k|k}$, though for notational simplicity this is not explicit in the notation.

Consider the set ${\bf X}_{k:k+1}$ of trajectories in time interval $k:k+1$. We decompose ${\bf X}_{k:k+1} = {\bf Y} \uplus {\bf V} \uplus {\bf B}$, where ${\bf Y}=\{Y_1,\dots,Y_{n_y}\}$ is a set of trajectories present at both time $k$ and $k+1$, ${\bf V}=\{V_1,\dots,V_{n_v}\}$ is a set of trajectories present at time $k$ but not present at time $k+1$, and ${\bf B}=\{B_1,\dots,B_{n_b}\}$ is a set of trajectories born at time $k+1$. It is met that ${\bf x}_k = \tau_k({\bf Y}\uplus {\bf V})$. We denote the first state and the second state (if it exists) of trajectory $X$ as $X^|$ and $X^{||}$, respectively. Given the multitarget density $f({\bf x}_k|{\bf z}^k)$, the multitrajectory density $\pi({\bf X}_{k:k+1}|{\bf z}^k)$ can be evaluated as \eqref{eq:tvb1}. Note that the backward kernel density \eqref{eq:tvb1} takes nonzero values only when $n_k$ and $n_y+n_v$ take the same value.

We write a realization ${\bf X}^+_{k+1:K}= {\bf S}_{k+1:K}\uplus{\bf D}_{k+2:K}$ of the multitrajectory smoothing density $\pi({\bf X}_{k+1:K}|{\bf z}^K)$ as the disjoint union of the set ${\bf S}_{k+1:K}$ of trajectories present at time $k + 1$ and the set ${\bf D}_{k+2:K}$ of trajectories only present at time $k + 2$ or afterwards. We also write ${\bf X}_{k:K} = {\bf U}_{k:K}\uplus{\bf D}^+_{k+2:K}$ as the disjoint union of the set ${\bf U}_{k:K}$ of trajectories present at time $k$ or time $k+1$ and the set ${\bf D}^+_{k+2:K}$ of trajectories only present at time $k + 2$ or afterwards. It is met that ${\bf U}_{k:k+1} = {\bf X}_{k:k+1} = {\bf Y} \uplus {\bf V} \uplus {\bf B}$ by construction. The backward kernel density can then be evaluated at ${\bf X}_{k:K}$ as \eqref{eq:ud1}.

We further write ${\bf U}_{k+1:K} = \{\bar{Y}_1,\dots,\bar{Y}_{n_y},\bar{B}_1,\dots,\bar{B}_{n_b}\}$ where trajectory $\bar{X}$ in time interval $k+1:K$ is an extension of trajectory $X$ in time interval $k:k+1$ and write ${\bf S}_{k+1:K} = \{X_1,\dots,X_{n_{k+1|K}}\}$ where $n_{k+1|K}$ is the number of trajectories present at time $k+1$. It should be noted that $X$ and $\bar{X}$ correspond to trajectories of the same target but in different time intervals. It is met that $\bar{Y}^| = Y^{||}$ and that $\bar{B}^| = B^{|}$. By expressing both $f_{l_{1:n_k}}(\cdot)$ and $\delta_{{\bf S}_{k+1:K}}(\cdot)$ as summations over permutations of elements, we can rewrite the backward kernel density as \eqref{eq:bwk}. 

Rearranging the factors in \eqref{eq:bwk} yields \eqref{eq:bwk2}. The summation over $\sigma_f$ can be interpreted as the sum over all possible associations between the Bernoulli components in the multi-Bernoulli component specified by $l_{1:n_k}$, i.e., \eqref{eq:mb01} and the trajectories in ${\bf X}_{k:k+1}$ present at time $k$. The summation over $\sigma_s$ can be interpreted as the sum over all possible associations between the trajectories in ${\bf S}_{k+1:K}$ and the trajectories in ${\bf U}_{k+1:K}$. Since we have that ${\bf X}_{k:k+1} = {\bf U}_{k:k+1}$ and that ${\bf U}_{k:k+1}$ and ${\bf U}_{k+1:K}$ represent the trajectories of the same set of objects but in different time intervals, there is an one-to-one mapping between the elements in ${\bf U}_{k+1:K}$ and the elements in ${\bf X}_{k:k+1}$. If a trajectory in ${\bf X}_{k:k+1}$ is not paired with any trajectory in ${\bf U}_{k+1:K}$, it is not present after time $k$. Therefore, the summation over $\sigma_s$ can also be interpreted as the sum over all possible associations between the trajectories in ${\bf S}_{k+1:K}$ and the trajectories in ${\bf X}_{k:k+1}$.

We proceed to introduce the following sets of single trajectory hypotheses:
\begin{subequations}
    \begin{align}
        \mathbb{Y}_{k}&=\{(\iota,j):\iota\in \{l_1,\dots,l_{n_k}\},j\in\{1,\dots,n_{k+1|K}\}\},\\
        \mathbb{V}_{k}&=\{(\iota,0):\iota\in \{l_1,\dots,l_{n_k}\}\},\\
        \mathbb{B}_{k}&=\{(0,j):j\in\{1,\dots,n_{k+1|K}\}\},
    \end{align}
\end{subequations}
where $\mathbb{Y}_{k}$ contains the hypotheses of a trajectory that is present at both time $k$ and $k+1$, $\mathbb{V}_{k}$ contains the hypotheses of a trajectory that is only present at time $k$, and $\mathbb{B}_{k}$ contains the hypotheses of a trajectory that is only present at time $k+1$. The different single trajectory hypotheses can be interpreted as: hypothesis $h=(\iota,j)\in\mathbb{Y}_{k}$ means a target has single target filtering density $f^{\iota}(x)$ at time $k$ and trajectory $X_j$ in time interval $k+1:K$; hypothesis $h=(\iota,0)\in\mathbb{V}_{k}$ means a target has single target filtering density $f^{\iota}(x)$ at time $k$ and it is not present after time $k$; hypothesis $h=(0,j)\in\mathbb{B}_{k}$ means a target is not present at time $k$ and its trajectory in time interval $k+1:K$ is $X_j$. 

We denote the global association hypothesis space given $l_{1:n_k}$ as
\begin{multline}
    \mathbb{A}^{ l_{1:n_k} }_{k} = \Big\{{\cal A}^{ l_{1:n_k} }_{k}= {\cal Y}_{k}\uplus{\cal V}_{k}\uplus{\cal B}_{k} \Big| {\cal Y}_{k}\subset\mathbb{Y}_{k}, {\cal V}_{k}\subseteq\mathbb{V}_{k}, \\{\cal B}_{k}\subseteq\mathbb{B}_{k}, |{\cal Y}_{k}|+|{\cal V}_{k}| = n_k, |{\cal Y}_{k}|+|{\cal B}_{k}| = n_{k+1|K} \Big\}.
\end{multline} 
We can observe that there is a one-to-one mapping between a global association hypothesis ${\cal A}^{ l_{1:n_k} }_{k}\in \mathbb{A}^{ l_{1:n_k} }_{k}$ and a pair of permutations $(\sigma_f,\sigma_s)$ in \eqref{eq:bwk2} where $\sigma_f\in\Sigma_{n_k}$ and $\sigma_s\in\Sigma_{n_{k+1|K}}$. 
Specifically, for a pair of permutations $(\sigma_f,\sigma_s)$, its corresponding global association hypothesis is given by ${\cal Y}_k\uplus{\cal V}_k\uplus{\cal B}_k$ with
\begin{subequations}
    \begin{align}
        {\cal Y}_k &= \{(l_{\sigma_f(1)},\sigma_s(1)),\dots,(l_{\sigma_f(n_y)},\sigma_s(n_y))\},\\
        {\cal B}_k &= \{(0,\sigma_s(n_y+1)),\dots,(0,\sigma_s(n_y+n_b))\},\\
        {\cal V}_k &= \{(l_{\sigma_f(n_y+1)},0),\dots,(l_{\sigma_f(n_y+n_v)},0)\}.
    \end{align}
\end{subequations}

We denote the trajectory under single trajectory hypothesis $h$ in time interval $k:k+1$ as $X_h$ and its extension in time interval $k+1:K$ as $\bar{X}_h$. We can rewrite the backward kernel density as the summation over all possible global association hypotheses for each $l_{1:n_k}$ as \eqref{eq:bwk3}. Denoting $X_j=(t_j,x_j^{1:i_j})$, \eqref{eq:bwk3} can be written as
\begin{multline}
    \label{eq:single_tra_hypo}
    \pi({\bf U}_{k:K}\uplus{\bf D}^+_{k+2:K}|{\bf S}_{k+1:K}\uplus{\bf D}_{k+2:K},{\bf z}^K) \propto \sum_{l_{1:n_k}}Q_{l_{1:n_k}}\\\times \sum_{{\cal A}^{ l_{1:n_k} }_{k}\in\mathbb{A}^{ l_{1:n_k} }_{k}}\prod_{h\in{\cal A}^{ l_{1:n_k} }_{k}}w^hp(X|h)\delta_{{\bf D}_{k+2:K}}({\bf D}^+_{k+2:K}),
\end{multline}
where 
\begin{subequations}
    \begin{align}
        w^h &= \begin{cases}
            \int p_{\iota}(x) g\left(x_j^1|x\right)p_S\left(x\right) dx & h\in\mathbb{Y}_k\\
            \int p_{\iota}(x)\left( 1-p_S\left(x\right) \right) dx & h\in\mathbb{V}_k\\
            \lambda^b(x_j^1) & h\in\mathbb{B}_k,
        \end{cases}\label{eq:single_tra_weight}\\
        p(X|h) &= \begin{cases}
            p^{\mathbb{Y}}(X|h) & h\in\mathbb{Y}_k\\
            p^{\mathbb{V}}(X|h) & h\in\mathbb{V}_k\\
            \delta_{X_j}(X) & h\in\mathbb{B}_k,
        \end{cases}\label{eq:single_tra_density}\\
        p^{\mathbb{Y}}((t,x^{1:i})|h) &= \delta_k[t]\frac{p_{\iota}(x^1) g\left(x_j^1|x^1\right)p_S\left(x^1\right)}{\int p_{\iota}(x) g\left(x_j^1|x\right)p_S\left(x\right) dx}\delta_{x_j^{1:i_j}}(x^{2:i}),\\
        p^{\mathbb{V}}((t,x^1)|h) &= \delta_k[t]\frac{p_{\iota}(x^1)\left( 1-p_S\left(x^1\right) \right)}{\int p_{\iota}(x)\left( 1-p_S\left(x\right) \right) dx}.
    \end{align}
\end{subequations}

The rationale behind \eqref{eq:single_tra_hypo} is that a single trajectory hypothesis density integrates to one, so we should divide the unnormalized densities in \eqref{eq:bwk3} by their corresponding integrals. We can also identify the weights of different single trajectory hypotheses in \eqref{eq:single_tra_hypo} as the normalizing factors being divided. We can further observe that the parameterization of the RHS of \eqref{eq:single_tra_hypo} is similar to an $\text{MBM}_{01}$ but with the difference that the weights $Q_{l_{1:n_k}}\prod_{h\in{\cal A}_k^{l_{1:n_k}}}w^h$ are unnormalized. By normalizing the weights, the backward kernel $\text{MBM}_{01}$ density can be expressed as \eqref{eq:mbm01_backward_kernel}.

Drawing a sample ${{\bf X}}_{k:K}$ from \eqref{eq:mbm01_backward_kernel} consists of three steps. We first sample a data association hypothesis ${\cal A}^{ l_{1:n_k} }_{k}$. Next, we sample from the corresponding single trajectory densities \eqref{eq:single_tra_density} to obtain ${{\bf U}}_{k:K}$. Then we append ${{\bf U}}_{k:K}$ to ${{\bf D}}^+_{k+2:K}$ to obtain ${{\bf X}}_{k:K}$.

\subsection{A tractable implementation based on ranked assignments}\label{section:ranked_assignments}

Performing sampling directly using \eqref{eq:mbm01_backward_kernel} is computationally intractable due to the unknown associations between ${{\bf S}}_{k+1:K}$ and $f({\bf x}_k|{\bf z}^{k})$. One strategy to reduce the sampling space is by truncating the terms in the summations in \eqref{eq:mbm01_backward_kernel}. More specifically, we first select the $\text{MB}_{01}$ components with the highest weights by solving a ranked assignments problem\footnote{An alternative approach is using Gibbs sampling to find $\text{MB}_{01}$ components with high weights \cite{gibbs}.} using Murty's algorithm \cite{murty}, and then we only draw samples from the truncated $\text{MBM}_{01}$ \cite{trackingbasedontrajectories}. We proceed to present an alternative parameterization of \eqref{eq:mbm01_backward_kernel} that facilitates the formulation of the ranked assignments problem, see \eqref{eq:murty_weight}. According to the weight representation in \eqref{eq:murty_weight}, we can construct the cost matrix of size $n_{k|k}\times (n_{k+1|K}+2n_{k|k})$ as 
\begin{subequations}\label{eq:cost_matrix}
    \begin{align}
        C &= -\begin{bmatrix}
            C_1 & C_2 & C_3
        \end{bmatrix},\\
        \label{eq:cost_c1}
        C_1 &= \begin{bmatrix}
            \ln \left(\frac{r^1w^{(1,1)}}{w^{(0,1)}}\right) &\dots & \ln \left(\frac{r^1w^{(1,n_{k+1|K})}}{w^{(0,n_{k+1|K})}}\right)\\
            \vdots & \ddots &\vdots \\
            \ln \left(\frac{r^{n_{k|k}}w^{({n_{k|k}},1)}}{w^{(0,1)}}\right) &\dots& \ln \left(\frac{r^{n_{k|k}}w^{({n_{k|k}},n_{k+1|K})}}{w^{(0,n_{k+1|K})}}\right)
        \end{bmatrix},\\
        C_2 &= \text{diag}_{-\infty}\left(\ln\left(r^1w^{(1,0)}\right),\dots,\ln\left(r^{n_{k|k}}w^{(n_{k|k},0)}\right)\right),\\
        C_3 &= \text{diag}_{-\infty}\left(\ln\left(1-r^1\right),\dots,\ln \left(1-r^{n_{k|k}}\right)\right),
    \end{align}
\end{subequations}
where entries of matrices $C_2$ and $C_3$ that are not on the diagonal are set to $-\infty$.

The selection of single trajectory hypotheses \eqref{eq:single_tra_hypo} to be included in each mixture component of \eqref{eq:mbm01_backward_kernel} can be written as an $n_{k|k}\times (n_{k+1|K}+2n_{k|k})$ assignments matrix $S$ consisting of 0 or 1 entries such that each row sums to one and each column sums to zero or one. Note that, if the $i$th ($1\leq i\leq n_{k+1|K}$) column sums to zero, $X^i$ is a newborn trajectory at time $k+1$, and that, if the $i$th ($i\geq n_{k|k}+n_{k+1|K}$) column sums to one, the $i-n_{k|k}-n_{k+1|K}$th Bernoulli component of $f({\bf x}_k|{\bf z}^{k})$ is not included in the $\text{MB}_{01}$ to be sampled. The $M$-best $\text{MB}_{01}$ components that minimizes $\text{tr}(S^{\textrm{T}}C)$ can be obtained using Murty's algorithm. Pseudo-code for backward simulation for sets of trajectories is given in Algorithm \ref{alg1}.

\begin{algorithm}[!t]
    \caption{Pseudo code for backward simulation for sets of trajectories} 
    \label{alg1} 
    \begin{algorithmic}[1] 
        \REQUIRE MB filtering densities $f({\bf x}_k|{\bf z}_{1:k})$ for $k=1,\dots,K$.
        \ENSURE Backward sets of trajectories $\{{{\bf X}}^i_{1:K}\}_{i=1}^{T}$.
        \STATE Sample $\{{{\bf x}}^i_K\}_{i=1}^T$ from $f({\bf x}_K|{\bf z}_{1:K})$ and construct them as $\{{{\bf X}}^i_{K:K}\}_{i=1}^T$.
        \FOR{$k = K-1,\dots,1$}
        \FOR{$i = 1,\dots,T$}
        \STATE Separate ${{\bf X}}^i_{k+1:K} $ as ${{\bf S}}^i_{k+1:K} \uplus {{\bf D}}^i_{k+2:K}$.
        \STATE Construct the cost matrix \eqref{eq:cost_matrix} using \eqref{eq:mbm01} and \eqref{eq:single_tra_weight}, see Section \ref{section:ranked_assignments}.
        \STATE Find the $M$-best $\text{MB}_{01}$ of \eqref{eq:murty_weight} using Murty's algorithm.
        \STATE Sample an $\text{MB}_{01}$ from the truncated $\text{MBM}_{01}$.
        \STATE Sample a set ${{\bf U}}_{k:K}$ of trajectories from the selected $\text{MB}_{01}$ using \eqref{eq:single_tra_density}, see Section \ref{section:multi-Bernoulli}.
        \STATE ${{\bf X}}^i_{k:K} = {{{\bf U}}^i_{k:K}}\uplus {{\bf D}}^i_{k+2:K}$.
        \ENDFOR
        \ENDFOR
\end{algorithmic}
\end{algorithm}


\subsection{Linear Gaussian implementation}

We present the expressions of the weights and densities of different single trajectory hypotheses when the dynamic model and the target state densities are linear and Gaussian. Let the transition density be $g(x|x^\prime) = {\cal N}(x;Fx^\prime,Q)$ where $F$ is a state transition matrix, $Q$ is the covariance matrix of the process noise. Assume that the $i$th Bernoulli component in $f({\bf x}_k|{\bf z}_{1:k})$ has existence-conditioned state density $p^i(x)={\cal N}(x;m^i_{k|k},P^i_{k|k})$, and that the Poisson birth intensity is a Gaussian mixture $\lambda^b(x) = \sum_{i=1}^{N_b}w_{b,i}{\cal N}(x;m^{b,i},P^{b,i})$. Assume also that the target survival probability $p_S(\cdot) = p_S$ is constant. Then the weights of single trajectory hypotheses \eqref{eq:single_tra_weight} can be expressed as 
\begin{equation}
    w(h) = \begin{cases}
        p_S{\cal N}({x}^j;Fm^{\iota}_{k|k},FP^{\iota}_{k|k}F^{\textrm{T}}+Q) & h=(\iota,j)\in\mathbb{Y}_{k}\\
        1 - p_S& h=(\iota,0)\in\mathbb{V}_{k}\\
        \sum_{i=1}^{N_b}w_{b,i}{\cal N}({x}^j;m^{b,i},P^{b,i}) & h=(0,j)\in\mathbb{B}_{k}
    \end{cases}
\end{equation}
We proceed to describe how to draw samples from \eqref{eq:single_tra_density}. For single trajectory hypotheses $(\iota,0)\in\mathbb{V}_{k}$, a trajectory sample has initial time $k$ and its state can be drawn from $x^1 \sim {\cal N}(m^{\iota}_{k|k},P^{\iota}_{k|k})$. For single trajectory hypotheses $(0,j)\in\mathbb{B}_{k}$, a trajectory sample has initial time $k+1$ and its state is ${x}_j$. For single trajectory hypotheses $(\iota,j)\in\mathbb{Y}_{k}$, a trajectory sample has initial time $k$, its state at time $k+1$ is ${x}_j$, and its state at time $k$ can be drawn from $x^1 \sim {\cal N}(\mu_k,M_k)$ with
\begin{subequations}\label{eq:G_SOT}
    \begin{align}
        \mu_k &= m^{\iota}_{k|k} + P^{\iota}_{k|k}F^{\textrm{T}}P_{k+1|k}^{-1}({x}_j-Fm^{\iota}_{k|k}),\\
        M_k &=P^{\iota}_{k|k}- P^{\iota}_{k|k}F^{\textrm{T}}P_{k+1|k}^{-1}FP^{\iota}_{k|k},\\
        P_{k+1|k} &= Q + FP^{\iota}_{k|k}F^{\textrm{T}}.
    \end{align}
\end{subequations}

To further reduce computational complexity, we can use ellipsoidal gating on $\tau_{k+1}({\bf X}^+_{k+1:K})$ to remove unlikely associations. More specifically, if the squared Mahalanobis distance between ${x}_j$ and the predicted density of $p^i(x)={\cal N}(x;m^i_{k|k},P^i_{k|k})$, i.e., 
\begin{equation}
    ({x}_j - Fm^i_{k|k})^{\textrm{T}}(FP^i_{k|k}F^{\textrm{T}}+Q)^{-1}({x}_j - Fm^i_{k|k}),
\end{equation}
is larger than a predefined threshold, we can set its corresponding entry $C_{i,j}$ in cost matrix \eqref{eq:cost_c1} to $-\infty$.

\section{Simulation results}

\begin{figure}[!t]
    \center
    \includegraphics[width=\columnwidth]{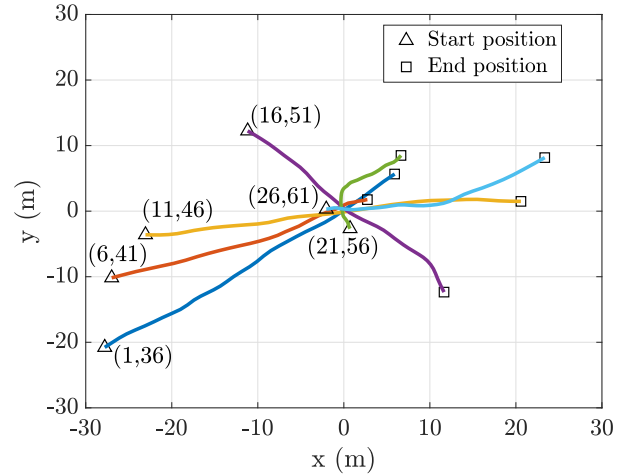}
    \caption{Six targets with different $(\text{birth time, \text{end time}})$ pairs move in close proximity around the midpoint.}
    \label{fig1}
\end{figure}

We evaluate the performance of the proposed multitrajectory smoother in a scenario with coalescence, see Fig. \ref{fig1}. Targets move following a constant velocity model. The process and measurement noises are all zero-mean Gaussian with standard deviation 0.1 for each dimension. The Poisson clutter rate is $30$ and the target detection probability is $0.7$. 

The (unlabelled) variational MB filter has shown excellent filtering performance when evaluated in scenarios with coalescence \cite{williams2014efficient,performanceevaluation}. Hence, we choose to apply the proposed smoothing algorithm on multitarget filtering densities obtained by a variational MB filter. In the variational MB filter, the posterior density at each time step is approximated as a Poisson MB using variational approximation, and the newborn targets are initiated from the Poisson intensity $\lambda_k^u(\cdot)$, typically a Gaussian mixture, representing undetected targets. Further, the recycling method of \cite{recycle} is applied to Bernoulli components with existence probability smaller than 0.1; they are approximated as being Poisson. Therefore, when running the proposed smoother backward, we need to compute the single trajectory hypothesis weight \eqref{eq:single_tra_weight} and density \eqref{eq:single_tra_density} using $\lambda_k^u(\cdot)$ instead of $\lambda^b(\cdot)$.

In the simulation, the gating size in probability is 0.999, the target survival probability is $p_S=0.97$, and the Poisson birth intensity $\lambda^b(\cdot)$ is a single Gaussian with weight $w_b=0.1$ centered at the midpoint with covariance covering the whole surveillance area. For the variational MB filter, Bernoulli components with existence probability smaller than $10^{-3}$ and Gaussian components in $\lambda^u_k(\cdot)$ with weights smaller than $10^{-3}$ are pruned. The set of targets estimate is formed by the means of the maximum a posteriori cardinality $n^{\star}$ Bernoulli components with highest existence probabilities. For the proposed smoother, 300 particles are used in backward simulation and Murty's algorithm is used to select at most 30 global hypotheses with highest weights. The set of trajectories estimate is selected as the particle with the highest global hypothesis weight accumulated over time. 

\begin{figure}[!t]
    \center
    \includegraphics[width=\columnwidth]{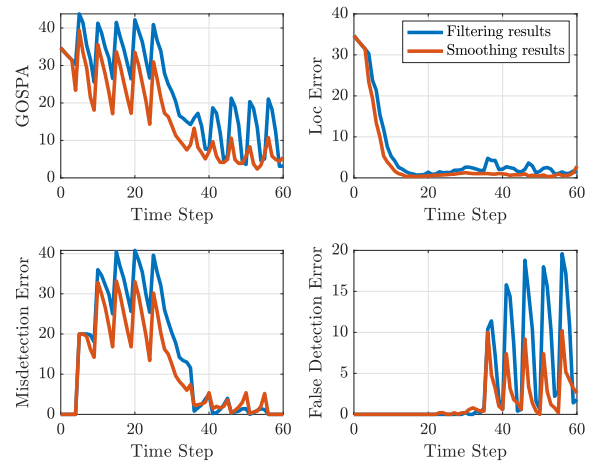}
    \caption{Performance evaluation using GOSPA metric.}
    \label{fig4}
\end{figure}

\begin{figure}[!t]
    \center
    \includegraphics[width=\columnwidth]{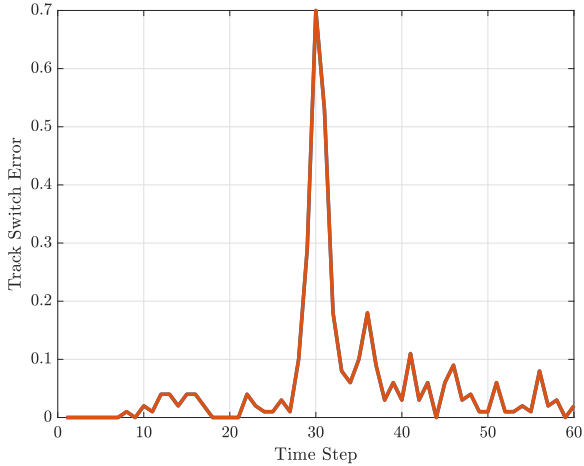}
    \caption{Track switch error evaluated using the trajectory metric \cite{trajectorymetric}.}
    \label{fig3}
\end{figure}

We evaluated the filtering and smoothing performance using GOSPA metric \cite{gospa} with parameters $\alpha=2,c=40,p=1$. The simulation results, averaged over 100 Monte Carlo runs, are presented in Fig. \ref{fig4}. Compared to the variational MB filter, the proposed multitrajectory smoother has improved localization performance, and in general it can detect target birth and death events more quickly. We also evaluated the tracking performance of the proposed smoother using the trajectory metric \cite{trajectorymetric} with parameters $\alpha=2,c=40,p=1$ and track switch penalty $\gamma=2$. Under this setting, the track switch error is equal to the average number of track switches. The average track switch error over time is shown in Fig. \ref{fig3}. It can be seen that the track switch error reaches its peak when the six targets are in close proximity, and in this worst case only about 0.7 track switch happens on average. Also, the average number of constructed trajectories is 6.79. These results show that the proposed smoother can build trajectories well based on (unlabelled) multitarget filtering densities.


\section{Conclusion}

We have presented the general backward-forward smoothing equation for sets of trajectories and proposed a tractable implementation of a multitrajectory smoother using backward simulation and ranked assignments. The effectiveness of the proposed approach is demonstrated in a simulation study.

\bibliographystyle{IEEEtran}
\bibliography{IEEEabrv,mybibfile}

\begin{figure*}[!t]
    \footnotesize
    \hrulefill
    \begin{align}
        g(Y|X) &= (1-|\tau_{\eta}(X)|)\delta_X(Y) + |\tau_{\eta}(X)|\Bigg(\big(1-p_S(y^{i^\prime})+p_S(y^{i^\prime})\delta_{\gamma-t^\prime+1}[i^\prime]\big)\delta_X\left((t^\prime,y^{1:i})\right)\prod_{j=i}^{i^\prime-1}g(y^{j+1}|y^{j})p_S(y^{j})\Bigg)\label{eq:multistep_transition}\\
        g(Y|X) &= \big(1-p_S(y^{i^\prime})+p_S(y^{i^\prime})\delta_{\gamma-t^\prime+1}[i^\prime]\big)\delta_X\left((t^\prime,y^{1:i})\right)\prod_{j=i}^{i^\prime-1}g(y^{j+1}|y^{j})p_S(y^{j})\label{eq:multistep_transition2}\\
        \pi_S({\bf Y}^\eta\uplus{\bf Z}^{\eta-1}|{\bf z}^{k}) &= \int \int \pi_g({\bf Y}^\eta\uplus{\bf Z}^{\eta-1}|{\bf D}\uplus{\bf A})\pi_{S^{-}}({\bf D}\uplus{\bf A}|{\bf z}^{k})\delta {\bf D} \delta {\bf A} = \int \pi_g({\bf Y}^\eta|{\bf A})\pi_{S^{-}}({\bf Z}^{\eta-1}\uplus{\bf A}|{\bf z}^{k})\delta {\bf A}\label{eq:set_integral_da}\\
        \pi_S({\bf Y}^\eta\uplus{\bf Z}^{\eta-1}|{\bf z}^{k}) &= \frac{1}{n!}\int \sum_{\sigma\in\Sigma_{n}}\prod_{j=1}^n g(Y_j|A_{\sigma(j)})\pi_{S^{-}}({\bf Z}^{\eta-1}\uplus\{A_1,\dots,A_{n}\}|{\bf z}^{k})d A_{1:n} \notag\\&= \int \prod_{j=1}^n g(Y_j|A_j)\pi_{S^{-}}({\bf Z}^{\eta-1}\uplus\{A_1,\dots,A_{n}\}|{\bf z}^{k})d A_{1:n}\label{eq:fb_sot_derivation_sub6}\\
        \pi_{\beta^{\eta+\iota}}({\bf B}^{\eta+\iota}) &= \frac{1}{n!}\int \sum_{\sigma\in\Sigma_{n}}\prod_{j=1}^n g(B_j|A_{\sigma(j)})\pi_{b^{\eta+\iota}}(\{A_1,\dots,A_{n}\})d A_{1:n} = \int \prod_{j=1}^n g(B_j|A_j)\pi_{b^{\eta+\iota}}(\{A_1,\dots,A_{n}\})d A_{1:n}\label{eq:fb_sot_derivation_sub7}\\
        \pi_{\beta^{\eta+\iota}}(\mathbf{B}^{\eta+\iota}) &= e^{-\int \lambda^b(x)dx}\prod_{(\eta+\iota,x^{1:i})\in{\bf B}^{\eta+\iota}}\lambda^b(x^1)\left( \left(1-p_S(x^i)+p_S(x^i)\delta_{\gamma-\eta-\iota+1}[i]\right) \prod_{j=1}^{i-1} g(x^{j+1} | x^j ) p_S(x^j) \right)\label{eq:fb_sot_derivation_sub9}
    \end{align}
    \hrulefill
\end{figure*}




\appendices

\section{Proof of Theorem \ref{theorem:mt_predict}}
\begin{proof}
    We start by presenting some preliminaries that are useful to the proof. We first clarify that if $(t,x^{1:i})\in{\bf W}^{\eta+1}$, then $\eta+1\leq t\leq \gamma$, $1\leq i\leq \gamma-\eta$; if $(t,x^{1:i})\in{\bf Y}^\eta$, then $\alpha \leq t \leq \eta$, $1\leq i \leq \gamma-\alpha+1$; and if $(t,x^{1:i})\in{\bf Z}^{\eta-1}$, then $\alpha \leq t \leq \eta-1$, $1\leq i \leq \eta-\alpha$. When no new trajectory is born, the number of trajectories in the set of all trajectories remains unchanged.\footnote{If a target dies, its trajectory remains, and therefore the number of trajectories is unchanged.} The multitrajectory transition density is
    \begin{equation}\label{eq:transition_no_birth}
        \pi_g(\{Y_1,\dots,Y_n\}|\{X_1,\dots,X_n\}) = \sum_{\sigma\in\Sigma_n}\prod_{j=1}^n g(Y_{\sigma(j)}|X_{j}),
    \end{equation}
    where $\Sigma_n$ is the set that includes all the permutations of $(1,\dots,n)$.
    
    Given single target trajectories $X=(t,x^{1:i})$ with $\alpha\leq t \leq t+i-1\leq\eta$ at time $\eta$ and $Y=(t^\prime,y^{1:i^\prime})$ with $\alpha \leq t^\prime \leq t^\prime+i^\prime-1\leq\gamma$ at time $\gamma$, the single trajectory transition density from $X$ to $Y$ is \eqref{eq:multistep_transition}. That is, if the trajectory has died before time $\eta$, the trajectory remains unaltered with probability one. If the trajectory exists at time $\eta+\iota$ with $0\leq \iota\leq \gamma-\eta-1$, it remains unaltered with probability $(1-p_S(\cdot))$ or the last target state is generated according to the single target transition density with probability $p_S(\cdot)$. We note that when $i^\prime=i$, the product of factors $\prod_{j=i}^{i^\prime-1}$ in \eqref{eq:multistep_transition} does not exist and therefore reduces to $1$, and in this case the trajectory dies at time $\eta$. Given $X=(t,x^{1:i})$, trajectory $Y$ must have the same initial time as $X$ and its length can vary from $i$ to $\gamma-t+1$. When the time step of the latest state of the trajectory is $\gamma$, i.e., $\delta_{\gamma-t^\prime+1}[i^\prime] = 1$, we no longer need to consider the possibility that the target will die at the next time step.
    
    Given single target trajectories $X=(t,x^{1:i})$ with $\eta+1 \leq t \leq t+i-1 \leq \gamma-1$ at sometime between $\eta+1$ and $\gamma-1$, and $Y=(t^\prime,y^{1:i^\prime})$ with $\eta+1\leq t^\prime\leq t^\prime+i^\prime-1 \leq \gamma$ at time $\gamma$, the single trajectory transition density from $X$ to $Y$ is \eqref{eq:multistep_transition2}, which can be considered a simplified version of \eqref{eq:multistep_transition} since it is known that trajectory $X$ exists at $\eta+1$ or afterwards.

    We use $\pi_S(\cdot)$ to denote the multitrajectory density at time $\eta+1$ for trajectories born before time $\eta+1$ and $\pi_{S^{-}}(\cdot)$ to denote multitrajectory density at time $\eta$. Given $\eta\geq k$ and that we only consider multitrajectory density conditioned on measurements up to time $k$, the set ${\bf W}^{\eta+1}$ of trajectories born at time $\eta+1$ and afterwards is independent of the set of ${\bf Y}^{\eta}\uplus{\bf Z}^{\eta-1}$ trajectories born before time $\eta+1$. This enables the use of the convolution formula to rewrite the set density of ${\bf W}^{\eta+1}\uplus{\bf Y}^\eta\uplus{\bf Z}^{\eta-1}$ as 
    \begin{equation}
        \pi({\bf X}_{\alpha:\gamma}|{\bf z}^{k}) = \sum_{{\bf A}\subseteq {\bf X}_{\alpha:\gamma}} \pi_{\beta}({\bf A})\pi_S({\bf W}^{\eta+1}\uplus{\bf Y}^\eta\uplus{\bf Z}^{\eta-1} \setminus {\bf A}|{\bf z}^{k}).
    \end{equation}
    As $\pi_{S}(\cdot)$ is the multitrajectory density for trajectories born before time $\eta+1$
    and $\pi_{\beta}$ is the multitrajectory density for trajectories born at time $\eta+1$ and afterwards, $\pi_S({\bf W}^{\eta+1}\uplus{\bf Y}^\eta\uplus{\bf Z}^{\eta-1} \setminus {\bf A}|{\bf z}^{k})$ is different from zero only if ${\bf W}^{\eta+1}\uplus{\bf Y}^\eta\uplus{\bf Z}^{\eta-1} \setminus {\bf A} \subseteq {\bf Y}^\eta\uplus {\bf Z}^{\eta-1}$, i.e., 
    ${\bf W}^{\eta+1} \subseteq {\bf A}$, and $\pi_{\beta}({\bf A})$ is different from zero only if ${\bf A} \subseteq {\bf W}^{\eta+1}$. Thus, we can conclude that $\pi({\bf X}_{\alpha:\gamma}|{\bf z}^{k})$ is different from zero only if ${\bf W}^{\eta+1} = {\bf A}$, which yields
    \begin{equation}
        \pi({\bf X}_{\alpha:\gamma}|{\bf z}^{k}) = \pi_S({\bf Y}^\eta\uplus{\bf Z}^{\eta-1}|{\bf z}^{k})\pi_{\beta}({\bf W}^{\eta+1}).
    \end{equation}
    In what follows, we prove \eqref{eq:fb_sot_derivation_sub3} and \eqref{eq:multistep_birth}.

    The multitrajectory density $\pi_S({\bf Y}^\eta\uplus{\bf Z}^{\eta-1}|{\bf z}^{k})$ is given by 
    \begin{equation}
        \pi_S({\bf Y}^\eta\uplus{\bf Z}^{\eta-1}|{\bf z}^{k}) = \int \pi_g({\bf Y}^\eta\uplus{\bf Z}^{\eta-1}|{\bf W}^\prime)\pi_{S^{-}}({\bf W}^\prime|{\bf z}^{k}) \delta {\bf W}^\prime.
    \end{equation}
    Partitioning ${\bf W}^\prime = {\bf D}\uplus{\bf A}$, where ${\bf D}$ and ${\bf A}$, respectively, represent dead and alive trajectories at time $\eta$, the set integral over ${\bf W}^\prime$ can be calculated as the set integral over ${\bf D}$ and ${\bf A}$, see \eqref{eq:set_integral_da}.
    Evaluating this expression for ${\bf Y}^\eta=\{Y_1,\dots,Y_n\}$ and using \eqref{eq:transition_no_birth} yields \eqref{eq:fb_sot_derivation_sub6}. The second equality of \eqref{eq:fb_sot_derivation_sub6} holds is because the permutation of $(1,\dots,n)$ does not affect the integral over the $A_{1:n}$. The proof of \eqref{eq:fb_sot_derivation_sub3} is finished by substituting \eqref{eq:multistep_transition} into \eqref{eq:fb_sot_derivation_sub6}. 

    Denote the set of trajectories born at time $\eta+\iota$ with $1\leq \iota\leq \gamma-\eta$ as ${\bf B}^{\eta+\iota}$ and its corresponding multitrajectory  density as $\pi_{\beta^{\eta+\iota}}(\cdot)$. We have that $\uplus_{\iota=1}^{\gamma-\iota}{\bf B}^{\eta+\iota} = {\bf W}^{\eta+1}$, and because trajectories born and evolve independently of each other, it holds that
    \begin{equation}
        \pi_\beta({\bf W}^{\eta+1}) = \sum_{\uplus_{\iota=1}^{\gamma-\eta}{\bf A}^{\eta+\iota} = {\bf W}^{\eta+1}}\prod_{\iota=1}^{\gamma-\eta}\pi_{\beta^{\eta+\iota}}({\bf A}^{\eta+\iota}).
    \end{equation}
    As $\pi_{\beta^{\eta+\iota}}(\cdot)$ is the multitrajectory density for trajectories born at time $\eta+\iota$, $\pi_{\beta^{\eta+\iota}}({\bf A}^{\eta+\iota})$ is different zero only if ${\bf A}^{\eta+\iota} = {\bf B}^{\eta+\iota}$. This yields 
    \begin{equation}\label{eq:fb_sot_derivation_sub8}
        \pi_\beta({\bf W}^{\eta+1}) = \prod_{\iota=1}^{\gamma-\eta}\pi_{\beta^{\eta+\iota}}({\bf B}^{\eta+\iota}).
    \end{equation}
    Denote the multitrajectory birth density at time $\eta+\iota$ as $\pi_{b^{\eta+\iota}}(\cdot)$, and for a Poisson birth model it has the expression 
    \begin{equation}\label{eq:Poissonbirth}
        \pi_{b^{\eta+\iota}}({\bf X}) = e^{-\int \lambda^b(x)dx}\prod_{(\eta+\iota,x^1)\in {\bf X}}\lambda^b(x^1).
    \end{equation}
    The multitrajectory density $\pi_{\beta^{\eta+\iota}}({\bf B}^{\eta+\iota})$ is
    \begin{equation}\label{eq:36}
        \pi_{\beta^{\eta+\iota}}({\bf B}^{\eta+\iota}) = \int \pi_g({\bf B}^{\eta+\iota} | {\bf A})\pi_{b^{\eta+\iota}}({\bf A}) \delta {\bf A}.
    \end{equation}
    Evaluating \eqref{eq:36} for ${\bf B}^{\eta+\iota}=\{B_1,\dots,B_n\}$ and using \eqref{eq:transition_no_birth} yields \eqref{eq:fb_sot_derivation_sub7}.
    Plugging \eqref{eq:multistep_transition2} and \eqref{eq:Poissonbirth} into \eqref{eq:fb_sot_derivation_sub7} yields \eqref{eq:fb_sot_derivation_sub9}. The proof of \eqref{eq:multistep_birth} is finished by substituting \eqref{eq:fb_sot_derivation_sub9} into \eqref{eq:fb_sot_derivation_sub8}. 

\end{proof}

\section{Proof of Corollary \ref{corollary:mt_prediction}}

\begin{proof}
    We observe that only the middle factor on the RHS of \eqref{eq:fb_sot_derivation_sub3} depends on $\alpha$. We also recall that $f(\cdot)$ is the multi-target predicted density which meets $\pi({\bf X}_{k+1:k+1}|{\bf z}^{k})=f({\bf x}_{k+1}|{\bf z}^{k})$. Therefore, setting $\eta = k+1$ and dividing the first factor on the RHS of \eqref{eq:fb_sot_derivation_sub3} from the LHS of \eqref{eq:fb_sot_derivation_sub3} yields 
    \begin{equation}\label{eq:corollary1_proof}
        \frac{\pi({\bf X}_{k:\gamma}|{\bf z}^{k})}{\pi({\bf X}_{k:k+1}|{\bf z}^{k})} = \frac{\pi({\bf X}_{k+1:\gamma}|{\bf z}^{k})}{f({\bf x}_{k+1}|{\bf z}^{k})},
    \end{equation}
    where we set $\alpha=k$ on the LHS and $\alpha = k+1$ on the RHS. By rearranging \eqref{eq:corollary1_proof}, we obtain \eqref{eq:corollary1}.
\end{proof}

\section{Proof of Theorem \ref{theorem:FB_SoT}}
\begin{proof}
    We denote ${\bf X}^+_{k+1:K}$ as a copy of the same variable of ${\bf X}_{k:K}$, restricted to a narrower time interval. Then the multitrajectory density of interest is
    \begin{equation}\label{eq:fb_sot_derivation}
        \begin{split}
            &\pi({\bf X}_{k:K}|{\bf z}^{K}) = \int \pi({\bf X}_{k:K},{\bf X}^+_{k+1:K}|{\bf z}^{K}) \delta {\bf X}^+_{k+1:K}\\
            &= \int \pi({\bf X}_{k:K}|{\bf X}^+_{k+1:K},{\bf z}^{k})\pi({\bf X}^+_{k+1:K}|{\bf z}^{K}) \delta {\bf X}^+_{k+1:K}\\
            &=\int \frac{\pi({\bf X}_{k:K}|{\bf z}^{k})\pi_g({\bf X}^+_{k+1:K}|{\bf X}_{k:K})}{\pi({\bf X}^+_{k+1:K}|{\bf z}^{k})} \pi({\bf X}^+_{k+1:K}|{\bf z}^{K}) \delta {\bf X}^+_{k+1:K}\\
            &=\int \frac{\pi({\bf X}_{k:K}|{\bf z}^{k}) \delta_{{\bf X}_{k+1:K}}({\bf X}^+_{k+1:K})}{\pi({\bf X}^+_{k+1:K}|{\bf z}^{k})}\pi({\bf X}^+_{k+1:K}|{\bf z}^{K}) \delta {\bf X}^+_{k+1:K}\\
            &= \frac{\pi({\bf X}_{k:K}|{\bf z}^{k})\pi({\bf X}_{k+1:K}|{\bf z}^{K})}{\pi({\bf X}_{k+1:K}|{\bf z}^{k})}.
        \end{split}
    \end{equation}
The first line follows the law of total probability. In the fourth line, we observe that $\pi_g({\bf X}^+_{k+1:K}|{\bf X}_{k:K})$ is a multitrajectory delta function. In the fifth line, we formulate a transition density from ${\bf X}_{k:K}$ to ${\bf X}_{k+1:K}$ using multitrajectory delta function $\delta_{{\bf X}_{k+1:K}}({\bf X}^+_{k+1:K})$ and the integral over ${\bf X}^+_{k+1:K}$ can be cancelled out by applying the prediction equation for sets of trajectories \cite[Eq. 8]{trackingbasedontrajectories}. Applying Corollary \ref{corollary:mt_prediction}, we have
    \begin{equation}\label{eq:fb_sot_derivation_sub1}
        \frac{\pi({\bf X}_{k:K}|{\bf z}^{k})}{\pi({\bf X}_{k+1:K}|{\bf z}^{k})} = \frac{\pi({\bf X}_{k:k+1}|{\bf z}^{k})}{f({\bf x}_{k+1}|{\bf z}^{k})}.
    \end{equation}
    The proof is finished by plugging \eqref{eq:fb_sot_derivation_sub1} into \eqref{eq:fb_sot_derivation}.
\end{proof}


\end{document}